\documentclass[
reprint,
superscriptaddress,
amsmath,amssymb,
aps,
prb,
]{revtex4-2}

\usepackage{movement-arrows}
\usepackage{physics}
\usepackage{graphicx}
\usepackage{dcolumn}
\usepackage{upgreek}
\usepackage{mathrsfs}
\usepackage{setspace}
\usepackage{bm}
\usepackage[breaklinks=true,colorlinks=true,linkcolor=blue,urlcolor=blue,citecolor=blue]{hyperref}
\allowdisplaybreaks[4]
\begin{document}
	
\preprint{APS/123-QED}
	
\title{Probing Hilbert space fragmentation with strongly interacting Rydberg atoms}

\author{Fan Yang}
\email{fanyangphys@gmail.com}
\affiliation{Niels Bohr Institute, University of Copenhagen, Blegdamsvej 17, 2100 Copenhagen, Denmark}

\author{Hadi Yarloo}
\affiliation{Department of Physics and Astronomy, Aarhus University, 8000 Aarhus C, Denmark}

\author{Hua-Chen Zhang}
\affiliation{Department of Physics and Astronomy, Aarhus University, 8000 Aarhus C, Denmark}


\author{Klaus M{\o}lmer}
\affiliation{Niels Bohr Institute, University of Copenhagen, Blegdamsvej 17, 2100 Copenhagen, Denmark}

\author{Anne E. B. Nielsen}
\affiliation{Department of Physics and Astronomy, Aarhus University, 8000 Aarhus C, Denmark}


\begin{abstract}
Hilbert space fragmentation provides a mechanism to break ergodicity in closed many-body systems. Here, we propose a feasible scheme to explore this exotic paradigm on a Rydberg quantum simulator. We show that the Rydberg Ising model in the large detuning regime can be mapped to a generalized folded XXZ model featuring a strongly fragmented Hilbert space. The emergent Hamiltonian, however, displays distinct time scales for the transport of a magnon and a hole excitation. This interesting property facilitates a continuous tuning of the Krylov-subspace ergodicity, from the integrable regime, to the Krylov-restricted thermal phase, and eventually to the statistical bubble localization region. By further introducing nonlocal interactions, we find that both the fragmentation behavior and the ergodicity of the Krylov subspace can be significantly enriched. We also examine the role of atomic position disorders and identify a symmetry-selective many-body localization transition. We demonstrate that these phenomena manifest themselves in quench dynamics, which can be readily probed in state-of-the-art Rydberg array setups.
\end{abstract}
	
\maketitle
\section{Introduction}
Statistical mechanics stands out as one of the most successful theories in physics, endowed with the ability to reproduce thermodynamics of macroscopic systems through modest assumptions \cite{landau2013statistical,callen1991thermodynamics,greiner2012thermodynamics,pathria2016statistical}. As a cornerstone of the theory, a microcanonical ensemble can describe an isolated, closed system based on the postulate of equal probabilities of microstates. At the microscopic level, a closed system is governed by quantum mechanics, where the quantum coherence precludes such an intrinsic ergodicity. In this case, thermal equilibrium can be locally reached by coherent interactions with the rest of the system which serves as an effective reservoir. The ergodic assumption of a quantum system is then reconciled by the eigenstate thermalization hypothesis (ETH) \cite{deutsch2018eigenstate}.

Despite the success of the ETH, recent studies have witnessed a number of paradigms to break the ergodicity of a closed system \cite{turner2018weak,alet2018many,abanin2019colloquium,moudgalya2022quantum,buca2023unified}. A prominent example is many-body localization (MBL) \cite{kjall2014many,vosk2015theory,schreiber2015observation,smith2016many}, in which strong, quenched disorders can hinder the thermalization of an interacting system. Recently, another paradigm has been proposed for the ergodicity breaking: the Hilbert space fragmentation (HSF), through which the Hilbert space fractures into exponentially many disjoint subspaces that cannot be distinguished by symmetry quantum numbers \cite{moudgalya2022quantum}. These Krylov subspaces can significantly slow down or even freeze the quantum dynamics, while suppressing the entanglement between different regions of the system. To date, a host of HSF models have been proposed, including the pair hopping model \cite{seidel2005incompressible,moudgalya2019thermalization,moudgalya2020quantum,kohlert2023exploring}, the spin-1 dipole-conserving model \cite{pai2019localization,sala2020ergodicity,rakovszky2020statistical,khemani2020localization}, and the folded XXZ model \cite{de2019dynamics,yang2020hilbert,zadnik2021foldedI,zadnik2021foldedII,pozsgay2021integrable,langlett2021hilbert}, etc. However, understanding the thermodynamic behavior of these models still constitutes a challenging task, as the required large Hilbert space dimension can make numerical simulations intractable.

Recent advances in Rydberg tweezer arrays have enabled unprecedented tunability of a many-body system comprised of individual atoms \cite{browaeys2020many,henriet2020quantum,morgado2021quantum,wu2021concise}, pushing quantum simulations closer to a regime inaccessible with classical computers \cite{ebadi2021quantum,scholl2021quantum,chen2023continuous,bluvstein2024logical,ma2023high,choi2023preparing,singh2023mid,PhysRevLett.130.243001,PhysRevX.13.041035}. This stimulates pioneering research on quantum thermalization in Rydberg arrays, including demonstration of the detailed balance principle \cite{kim2018detailed} and observation of quantum many-body scars \cite{bernien2017probing,bluvstein2021controlling}. Developing realistic protocols on Rydberg platforms to explore the HSF can thus greatly advance the study of this unique ergodicity-breaking paradigm.

Despite its significance, it is hard to construct a spin model supporting diverse HSF behaviors, as the required multibody interactions often come from high-order perturbations \cite{chen2021emergent}, which result in a slow dynamics severely limited by decoherence of the system. In this article, we develop a feasible scheme for studying the HSF in a Rydberg quantum simulator. We show that with a detuned laser driving and a two-body Rydberg interaction, one can effectively engineer a four-body interaction by just the second-order perturbation, which maps the system onto a spin model featuring a nontrivially fragmented Hilbert space. We show that both the fragmentation structure and the ergodicity of typical Krylov subspaces can be flexibly tuned by adjusting the strength and the range of Rydberg interactions. The interplay between kinetic constraints and atomic position disorders furthermore gives rise to a symmetry-sector selective MBL transition, in which a strong disorder can localize the motion of the holes while leaving the magnons thermalized. All these interesting phenomena are characterized by a detailed study of the level-spacing statistics, eigenstate properties, as well as dynamical signatures. The feasibility of our scheme is highlighted by a recent experiment \cite{kim2023realization}, which demonstrates constrained dynamics in the context of few-body bound states.

The paper is organized as follows. In Sec.~\ref{sec:sec2}, we formulate an effective model for describing the emergent Hilbert space fragmentation in a Rydberg Ising chain subjected to nearest-neighbor interactions. The formalism is then generalized to the realistic regime with incorporation of nonlocal Rydberg interactions (Sec.~\ref{sec:sec3}). In Sec.~\ref{sec:sec4}, we investigate the ergodicity of the system in the presence of atomic position disorders and identify a possible ergodic-MBL transition for certain symmetry sectors. In Sec.~\ref{sec:sec5}, we develop measurement protocols for observing signatures of the fragmentation and the localization in a Rydberg quantum simulator. We conclude with several prospects for further study in Sec.~\ref{sec:sec6}.

\section{Hilbert space fragmentation in a Rydberg Ising chain}\label{sec:sec2}
\begin{figure}[b]
	\centering
	\includegraphics[width=\linewidth]{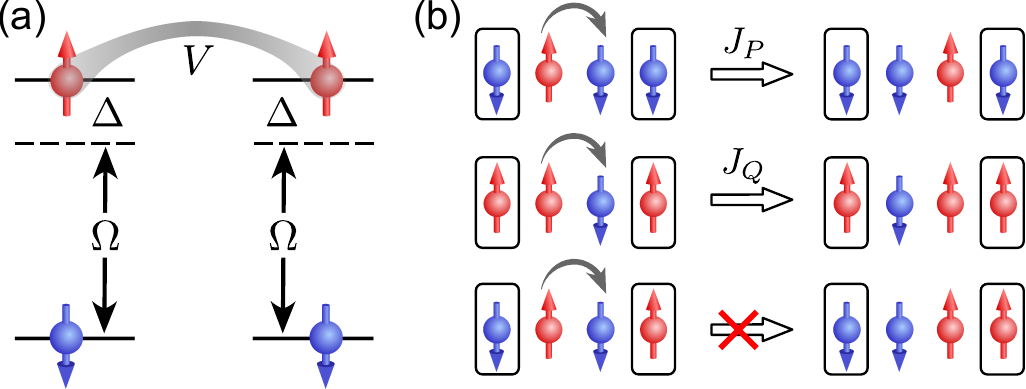}
	\caption{(a) Level structure for two individual atoms. A laser with Rabi frequency $\Omega$ and detuning $\Delta$ drives the ground state (encoded as the spin-down state $|{\downarrow}\rangle$) to the Rydberg state (encoded as the spin-up state $|{\uparrow}\rangle$). If the atoms are close to each other and both atoms are in the Rydberg state, there will be an extra contribution $V$ to the energy. (b) Illustration of the constrained flip-flop terms, where the hopping of a magnon and a hole has distinct rates $J_P$ and $J_Q$. Conservation of the domain wall number forbids the spin exchange sandwiched between atoms in opposite spin states.} \label{fig:fig1}
\end{figure}

The system under consideration is a one-dimensional atom array with $L$ atoms. Each individual atom can be represented by a spin-1/2 particle, where the spin flip between the ground state $\ket{\circ}=\ket{\downarrow}$ and the Rydberg state $\ket{\bullet}=\ket{\uparrow}$ is coherently driven by a laser field. The van der Waals (vdW) interaction between Rydberg states induces a state-dependent level shift [see Fig.~\ref{fig:fig1}(a)]. Considering only nearest-neighbor (NN) interaction, the Hamiltonian of the system is equivalent to an Ising model with open boundaries
\begin{equation}
	{H}_\mathrm{exact} = \frac{\Omega}{2}\sum_{i=1}^{L}{\sigma}_i^x +  \Delta \sum_{i=1}^{L}  {n}_i + V\sum_{i=1}^{L-1}  {n}_i {n}_{i+1} \label{eq:eq1},
\end{equation}
where $\sigma_i^x=|{\uparrow}\rangle_i \langle {\downarrow}|+|{\downarrow}\rangle_i \langle {\uparrow}|$ describes a single-spin flip, and $n_i=|{\uparrow}\rangle_i \langle {\uparrow}|=(\sigma_i^z+\mathbb{I})/2$ denotes the local Rydberg population. The Rabi frequency and the detuning of the laser driving are denoted by $\Omega$ and $\Delta$, respectively, while the NN interaction is assumed to be repulsive ($V>0$). While the NN model is an ideal model, we show in Sec.~\ref{sec:sec3} that a weak long-range interaction tail will not affect the state connectivity in the short-time dynamics.

\subsection{Effective Hamiltonian and emergent kinetic constraints}\label{subsec:sec2a}
Strong interactions can impose exotic kinetic constraints on the many-body dynamics. In a resonantly driven ($\Delta=0$) Rydberg chain with strong interactions ($V\gg\Omega$), the kinetic constraint manifests itself as the well-known Rydberg blockade, which prohibits simultaneous Rydberg excitations of neighboring sites, as described by the PXP model $H_\mathrm{PXP}=\sum_i \mathcal{P}_{i-1}\sigma_i^x \mathcal{P}_{i+1}$ with $\mathcal{P}_i=1-n_i$ the ground-state projector. Here, we shall focus on a different regime, where the laser driving is far-off resonant ($\Delta\gg\Omega$). In this dressing regime, the laser perturbatively couples the ground state to the Rydberg state. As a result, a single atom cannot be independently excited due to the large energy offset, but its state can flip via a virtual spin-exchange process that conserves the total number of Rydberg excitations $n_\mathrm{R}=\sum_i n_i$ \cite{yang2019quantum,kim2023realization}, or equivalently, the total magnetization $S_z=\sum_i\sigma_i^z$.

In order to describe the effective dynamics governed by the above U(1) symmetry, we carry out the Schrieffer-Wolff (SW) transformation of the original model \cite{bravyi2011schrieffer}. To this end, we first decompose the exact Hamiltonian into ${H}_\mathrm{exact}={H}_0+{\Omega}_D$, where
\begin{equation}
	{H}_\mathrm{0} =  \Delta \sum_{i=1}^{L} n_i+ V\sum_{i=1}^{L-1} {n}_i {n}_{i+1}\quad \mathrm{and}  \quad {\Omega}_D=\frac{\Omega}{2}\sum_{i=1}^{L}{\sigma}_i^x \label{eq:eq2}
\end{equation}
denote the diagonal part of the Hamiltonian and the off-diagonal perturbation, respectively. In the SW transformation, the Hamiltonian is transformed into $H_\mathrm{eff}=e^{\mathcal{S}}H_\mathrm{exact}e^{-\mathcal{S}}=H_0+H_\mathrm{eff}^{(1)}+H_\mathrm{eff}^{(2)}+\cdots$, where the anti-Hermitian generator $\mathcal{S}$ can be expanded in orders of the dressing parameter $\Omega/\Delta$, and $H_\mathrm{eff}^{(n)}$ denotes the $n$-th order effective Hamiltonian. Then, by choosing the generator $\mathcal{S}$ satisfying $[\mathcal{S},H_0]+\Omega_D=0$, the leading effective Hamiltonian appears at the second order
\begin{equation}
	H_\mathrm{eff}^{(2)}=[\mathcal{S},\Omega_D]/2.\label{eq:eq3}
\end{equation}
Since $n_\mathrm{R}$ is almost conserved, the effective Hamiltonian is approximately given by  $H_\mathrm{eff}={\Pi}H_\mathrm{eff}^{(2)}{\Pi}$, where ${\Pi}$ projects on terms that conserve the U(1) symmetry. In our system, the vdW interaction leads to state-dependent energy differences between neighboring blocks of $H_0$ labeled by $n_\mathrm{R}$. Therefore, to construct the generator, we need to project the neighboring sites of a single spin to all possible configurations, i.e.,
\begin{align}
	{\mathcal{S}} = \mathrm{i} \frac{\Omega}{2}\sum_{j=1}^{L} &\left( \frac{\mathcal{P}_{j-1}\sigma_j^y\mathcal{P}_{j+1}}{\Delta}+ 
	\frac{\mathcal{P}_{j-1}\sigma_j^y\mathcal{Q}_{j+1}}{\Delta+V}\right.\nonumber\\
	 &\left. +\,  \frac{\mathcal{Q}_{j-1}\sigma_j^y\mathcal{P}_{j+1}}{\Delta+V}  + 
	\frac{\mathcal{Q}_{j-1}\sigma_j^y\mathcal{Q}_{j+1}}{\Delta+2V}\right),
	\label{eq:eq4}
\end{align}
where $\mathcal{P}_{j}=|{\downarrow}\rangle_j \langle {\downarrow}|=1-n_j$ and $\mathcal{Q}_{j}=|{\uparrow}\rangle_j \langle {\uparrow}|=n_j$ denote the ground and the Rydberg state projector, respectively. To account for the edge effect under the open boundary condition, two virtual sites at each end ($\ket{\downarrow}_0$ and $\ket{\downarrow}_{L+1}$) are included to complete the above expression. With this explicit form of the generator, we can straightforwardly obtain the second-order effective Hamiltonian, whose off-diagonal term is a constrained XY model
\begin{align}
	\mathcal{H}_\mathrm{off} &= J_P\sum_{i=1}^{L-1} \mathcal{P}_{i-1} \left(\sigma_{i}^+ \sigma_{i+1}^- + \sigma_{i}^-\sigma_{i+1}^+\right)\mathcal{P}_{i+2} \nonumber \\
	& + J_Q\sum_{i=1}^{L-1}\mathcal{Q}_{i-1} \left(\sigma_{i}^+ \sigma_{i+1}^- + \sigma_{i}^-\sigma_{i+1}^+\right)\mathcal{Q}_{i+2} \nonumber \\
	& +  \frac{J_P+J_Q}{2} \sum_{i=1}^{L-1}\mathcal{P}_{i-1} \left(\sigma_{i}^+ \sigma_{i+1}^- + \sigma_{i}^-\sigma_{i+1}^+\right)\mathcal{Q}_{i+2}\nonumber \\
	& + \frac{J_P+J_Q}{2}\sum_{i=1}^{L-1} \mathcal{Q}_{i-1} \left(\sigma_{i}^+ \sigma_{i+1}^- + \sigma_{i}^-\sigma_{i+1}^+\right)\mathcal{P}_{i+2}.
	\label{eq:eq5}
\end{align}
where the induced spin-exchange interaction
\begin{equation}
	J_P = \frac{\Omega^2V}{4\Delta(\Delta+V)},\quad
	J_Q = \frac{\Omega^2V}{4(\Delta+V)(\Delta+2V)},\label{eq:eq6}
\end{equation}
is associated with the hopping of a magnon excitation $|\cdots{\circ}{\bullet}{\circ}\cdots\rangle$ and a hole excitation $|\cdots{\bullet}{\circ}{\bullet}\cdots\rangle$, respectively. It is important to notice that the original interaction strength $V$ is (at least) two orders of magnitude larger than the perturbative hopping strength $J_P$ and $J_Q$. As a consequence, only the first two terms (PXYP and QXYQ) in Eq.~(\ref{eq:eq5}) are dynamically relevant, while the last two terms (PXYQ and QXYP) are strongly suppressed due to the large energy offset $V$ [see Fig.~\ref{fig:fig1}(b)]. 

Such a constrained dynamics can be attributed to an emergent symmetry, i.e., the zeroth-order Hamiltonian $H_0=\Delta n_\mathrm{R}+ V n_\mathrm{NN}$ not only imposes a U(1) symmetry about $n_\mathrm{R}$, but also a symmetry with respect to conservation of the number of Rydberg dimers $n_\mathrm{NN}=\sum_i n_{i}n_{i+1}$, or equivalently, the number of domain walls ${D}_z=\sum_i(1-\sigma_i^z\sigma_{i+1}^z)/2$. Symmetry-breaking terms PXYQ and QXYP are thus prohibited. With the kinetic constraint and additional diagonal terms, the total effective Hamiltonian (dropping the constant terms) reads
\begin{align}
	H_\mathrm{eff} &= J_P\sum_{i=1}^{L-1}\mathcal{P}_{i-1} \left(\sigma_{i}^+ \sigma_{i+1}^- + \sigma_{i}^-\sigma_{i+1}^+\right)\mathcal{P}_{i+2} \nonumber \\
	& + J_Q\sum_{i=1}^{L-1}\mathcal{Q}_{i-1} \left(\sigma_{i}^+ \sigma_{i+1}^- + \sigma_{i}^-\sigma_{i+1}^+\right)\mathcal{Q}_{i+2} \nonumber \\
	&  +  \sum_{i=1}^{L}\mu_i {n}_i + U \sum_{i=1}^{L-1} {n}_i {n}_{i+1} + I \sum_{i=1}^{L} \mathcal{Q}_{i-1}{\sigma}_{i}^{z}\mathcal{Q}_{i+1},\label{eq:eq7}
\end{align}
where $\mu_1=\mu_{L}=\Delta+\Omega^2/2\Delta+J_P$, $\mu_i=\mu_1+J_P$ for $1<i<L$, $U=V-4J_P$, and $I = J_P-J_Q$ is the strength of the three-body interaction.

\begin{figure}
	\centering
	\includegraphics[width=\linewidth]{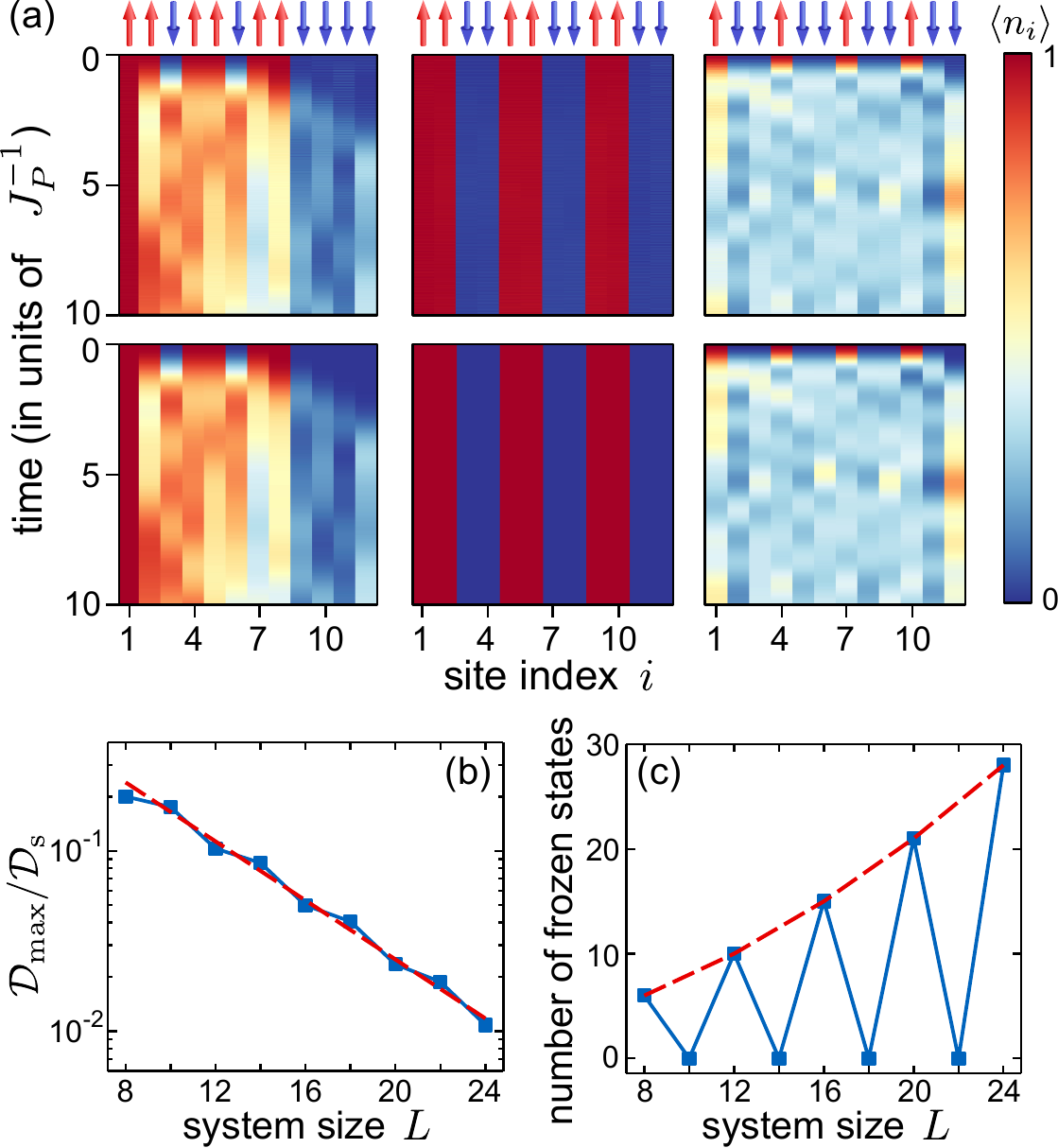}
	\caption{Characterization of the Hilbert space fragmentation. (a) Quench dynamics for different indicated initial states in a Rydberg chain with NN interactions. The upper and the lower panels show the evolution of the local Rydberg density $\langle n_{i} \rangle$ governed by the exact and the effective Hamiltonian, respectively. The parameters are $\Delta/\Omega=5$ and $V/\Delta=0.5$. (b) The largest symmetry sector exhibits strong HSF with $\mathcal{D}_\mathrm{max}/\mathcal{D}_\mathrm{s}\approx 1.08\times 0.828^L$ based on data up to $L=24$. The dashed line represents an exponential fit to the data. (c) The largest symmetry sector has $L^2/32+3L/8+1$ (dashed lines) frozen states when $L/2$ is even.}
	\label{fig:fig2}
\end{figure}
The above effective Hamiltonian is a generalized version of the folded XXZ model \cite{de2019dynamics,yang2020hilbert} that can exhibit HSF, i.e., the Hilbert space fractures into exponentially many dynamically disconnected subspaces even after resolving the symmetry quantum numbers $n_\mathrm{R}$ and $n_\mathrm{NN}$. The characteristic of the HSF is illustrated in Fig.~\ref{fig:fig2}(a), where we track evolution of the local Rydberg density $\langle n_i\rangle$ in a quench dynamics with different initial spin configurations. The left panel considers an initial state $\ket{{\bullet}{\bullet}{\circ}{\bullet}{\bullet}{\circ}{\bullet}{\bullet}{\circ}{\circ}{\circ}{\circ}}$ with $n_\mathrm{R}=6$ and $n_\mathrm{NN}=3$ in a system of $L=12$, where the dynamics is triggered by hopping of the holes. The initial state $\ket{{\bullet}{\bullet}{\circ}{\circ}{\bullet}{\bullet}{\circ}{\circ}{\bullet}{\bullet}{\circ}{\circ}}$ considered in the middle panel shares the same symmetry quantum numbers, but instead exhibits a completely frozen dynamics. Therefore, these two initial states should belong to disconnected subspaces that cannot be distinguished by conventional local symmetries. The quench dynamics in the right panel is initialized with $\ket{{\bullet}{\circ}{\circ}{\bullet}{\circ}{\circ}{\bullet}{\circ}{\circ}{\bullet}{\circ}{\circ}}$ and occurs on a shorter time scale compared with the left panel. This is because the corresponding symmetry sector $n_\mathrm{R}=4$ and $n_\mathrm{NN}=0$ is composed of states fully connected to each other through the faster magnon hopping ($J_P>J_Q$). As the evolution governed by the exact Hamiltonian (the upper panels) is nicely captured by the effective one (the lower panels), we will focus on the effective model in the rest of this article.

\begin{figure*}
	\centering
	\includegraphics[width=\linewidth]{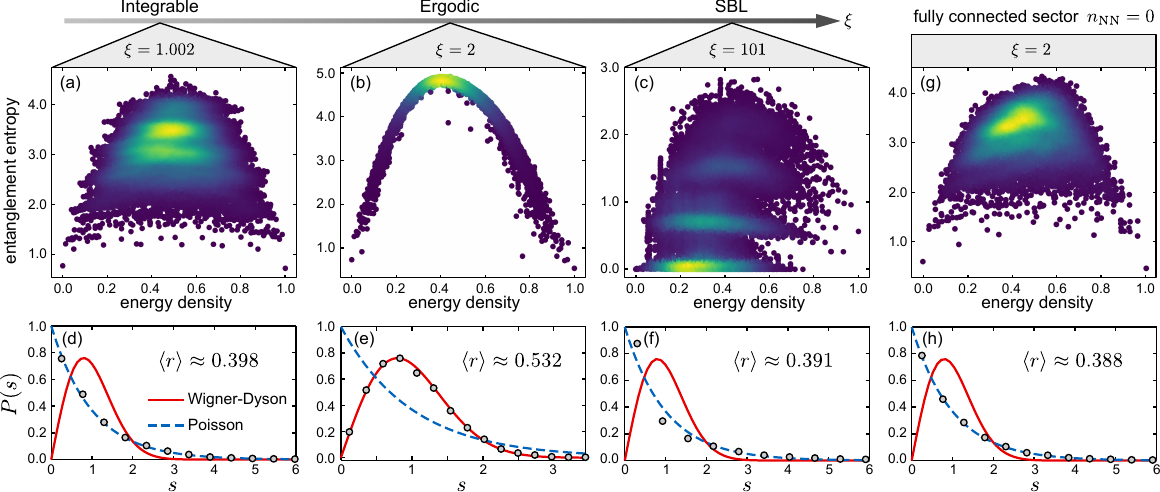}
	\caption{Half-chain von Neumann entanglement entropy of eigenstates and level statistics. (a)-(c) and (d)-(f) show the results for the largest Krylov subspace of the largest symmetry sector ($L=26$, $n_\mathrm{R}=13$, $n_\mathrm{NN}=6$) with indicated hopping ratio $\xi$. The subspace is generated by the root state $|{\bullet}{\bullet}{\circ} {\bullet}{\bullet}{\circ} {\bullet}{\bullet}{\circ} {\bullet}{\bullet}{\circ} {\bullet}{\bullet}{\circ} {\bullet}{\bullet}{\circ}
		{\bullet}{\circ} {\circ}{\circ}{\circ}{\circ}{\circ}{\circ}\rangle$ and has dimension $\mathcal{D}_\mathrm{max}=27132$. (g) and (h) show the results for the fully connected sector ($L=24$, $n_\mathrm{R}=8$, $n_\mathrm{NN}=0$) generated by the root state $|{\bullet}{\circ} {\circ} {\bullet}{\circ} {\circ} {\bullet}{\circ} {\circ} {\bullet}{\circ} {\circ} {\bullet}{\circ} {\circ} {\bullet}{\circ} {\circ} {\bullet}{\circ} {\circ} {\bullet}{\circ} {\circ}\rangle$ with $\xi=2$. Here, we consider the inversion-symmetric sector, which contains $12190$ states. The energy density for eigenenergy $E_n$ is defined as $\epsilon_n=(E_n-E_\mathrm{min})/(E_\mathrm{max}-E_\mathrm{min})$ with $E_\mathrm{max}$ ($E_\mathrm{min}$) the maximum (minimum) eigenenergy. The color scale in (a-c) and (g) represents the density of data points, with brighter colors indicating higher densities.}
	\label{fig:fig3}
\end{figure*}

To summarize, depending on the symmetry quantum numbers, the Hilbert space under the action of $H_\mathrm{eff}$ can be either fragmented or fully connected. We will mainly consider three typical symmetry sectors. (i) The largest symmetry sector $\{n_\mathrm{R}=L/2,n_\mathrm{NN}=L/4\}$ (for $L/2$ even) or $\{n_\mathrm{R}=L/2,n_\mathrm{NN}=L/4-1/2\}$ (for $L/2$ odd) forms a strongly fragmented space, e.g., the left panel of Fig.~\ref{fig:fig2}(a), in which the dimension of the largest subspace ($\mathcal{D}_\mathrm{max}$) is exponentially smaller than the size of the entire sector ($\mathcal{D}_\mathrm{s}$) with $\mathcal{D}_\mathrm{max}/\mathcal{D}_\mathrm{s}\sim 0.828^L$ [see Fig.~\ref{fig:fig2}(b)]. This sector also contains polynomially many ($\sim L^2$) frozen states when $L/2$ is even [see Fig.~\ref{fig:fig2}(c)], e.g., the middle panel of Fig.~\ref{fig:fig2}(a). (ii) The magnon sector with $n_\mathrm{NN}=0$ and $n_\mathrm{R}\in\{0,1,2,\ldots, (L+2)/3\}$, e.g., the right panel of Fig.~\ref{fig:fig2}(a), is a fully connected sector, where magnon excitations are separated from each other at a distance larger than one lattice site. (iii) The fully connected sectors formed solely by the hole excitation, which can be obtained by applying the spin-flip operation $\Pi_i\sigma_i^x$ on states in the magnon sector.

\subsection{Ergodicity of the Krylov subspace}\label{subsec:sec2b}
With the strong HSF, the system strongly violates the ETH in terms of the symmetry-resolved Hilbert space, e.g., the largest symmetry sector mentioned above. However, if we restrict ourselves to a specific Krylov subspace $\mathcal{K}_i=\mathrm{span}\{\ket{\psi_i},H_\mathrm{eff}\ket{\psi_i},H_\mathrm{eff}^2\ket{\psi_i},\cdots\}$ generated by a root state $\ket{\psi_i}$, the dynamics within $\mathcal{K}_i$ can otherwise be ergodic, a phenomenon referred to as the Krylov-restricted thermalization.

We now examine the ergodicity of the largest Krylov subspace of the largest symmetry sector of a given (even) $L$. The subspace can be generated by the root state
\begin{equation}
\ket{\mathrm{root}} = |\underbrace{\boxed{{\bullet}{\bullet}{\circ}} \boxed{{\bullet}{\bullet}{\circ}} \cdots \boxed{{\bullet}{\bullet}{\circ}}}_{3m}\ \underbrace{\boxed{{\circ}{\circ}\cdots{\circ}}}_{m}\rangle, \label{eq:eq8}
\end{equation}
for $L=4m$ and 
\begin{equation}
	\ket{\mathrm{root}} = |\underbrace{\boxed{{\bullet}{\bullet}{\circ}} \boxed{{\bullet}{\bullet}{\circ}} \cdots \boxed{{\bullet}{\bullet}{\circ}}}_{3m} {\bullet}{\circ} \underbrace{\boxed{{\circ}{\circ}\cdots{\circ}}}_{m}\rangle, \label{eq:eq9}
\end{equation}
for $L=4m+2$, which consist of small-size Rydberg clusters. These states are easier to prepare than the root states considered in previous works with a large magnon cluster $|\cdots{\bullet}{\bullet}{\bullet}{\bullet}\cdots\rangle$, as formation of big Rydberg clusters is strongly suppressed by the blockade effect. The major difference between the standard folded XXZ model and our generalized version lies in the asymmetric hopping between a magnon and a hole in Eq.~\eqref{eq:eq7}. We thus choose the hopping ratio as a tunable parameter:
\begin{equation}
	\xi=\frac{\textrm{magnon-hopping\ strength}}{\textrm{hole-hopping\ strength}}=\frac{J_P}{J_Q}=1+\frac{2V}{\Delta},
\end{equation}
which can be tuned from unity to infinity by increasing the interaction strength $V$.

To characterize the ergodicity of the considered Krylov subspace, we calculate the half-chain von Neumann entanglement entropy of eigenstates for a system size $L=26$. First, in the weak interaction limit $V/\Delta\rightarrow 0$ ($\xi\approx1$), we find that the entanglement entropy of the eigenstates significantly spreads, as opposed to the expectation from an ergodic phase fulfilling the ETH [see Fig.~\ref{fig:fig3}(a)]. In this specific regime, the hopping of the magnon and the hole is almost symmetric ($J_P\approx J_Q$), which also results in a vanishing three-body interaction $I=J_P-J_Q\approx0$ in Eq.~\eqref{eq:eq7}. With these approximations, the effective Hamiltonian is reduced to the ideal folded XXZ model, which is fully integrable and Bethe Ansatz solvable \cite{pozsgay2021integrable,gombor2021}. The approximate integrability is further confirmed by the study of level spacing statistics shown in Fig.~\ref{fig:fig3}(d). Here, we find that the distribution $P(s)$ of the normalized level spacing $s$ is close to the one describing an integrable system: the Poisson distribution $e^{-s}$ characterized by the mean level-spacing ratio $\langle r\rangle_\mathrm{Poisson}\approx0.386$, which is the average value of the ratio $r_n=\mathrm{min}\{s_n,s_{n+1}\}/\mathrm{max}\{s_n,s_{n+1}\}$ with $s_n=E_{n+1}-E_{n}$ the gap between adjacent eigenenergies $E_n$ and $E_{n+1}$.

The situation becomes drastically different when entering the intermediate interaction regime $V\sim\Delta$, in which both the asymmetric hopping and the three-body interaction become prominent. As shown in Fig.~\ref{fig:fig3}(b), the eigenstate entanglement entropy now forms a narrow ETH-like band, indicative of a broken integrability. The resulting level repulsion makes $P(s)$ approach the Wigner-Dyson distribution $(\pi/2)se^{-\pi s^2/4}$ that describes a Gaussian orthogonal ensemble (GOE) with $\langle r\rangle_\mathrm{GOE}\approx0.53$. In such an ergodic phase, the initial product state is expected to exhibit Krylov-restricted thermalization, which can be probed experimentally (see Sec.~\ref{sec:sec5}).

The ergodicity can be further modified if the interaction strength is increased to the limit $V/\Delta\rightarrow \infty$ ($\xi\gg1$), where the magnon hopping rate saturates to a finite value $\Omega^2/(4\Delta)$, while the hole becomes nearly frozen $J_Q\approx0$. The strong interaction limit thus falls into the statistical bubble localization (SBL) regime \cite{li2017statistical,yarloo2019emergent}, i.e., clusters of Rydberg excitations become insulating blocks, between which the isolated magnon excitations form an active bubble. The consequent ``bubble-neck'' structure of the eigenstates implies that the Krylov subspace fractures into multiple subsectors. The lack of level repulsion between these subsectors yields a mean level spacing ratio $\langle r\rangle\approx0.391$ close to the Poisson value again [see Fig.~\ref{fig:fig3}(f)], and the entanglement entropy $S(E)$ is no longer a smooth function of the eigenenergy $E$ [see Fig.~\ref{fig:fig3}(c)]. 

\begin{figure}
	\centering
	\includegraphics[width=\linewidth]{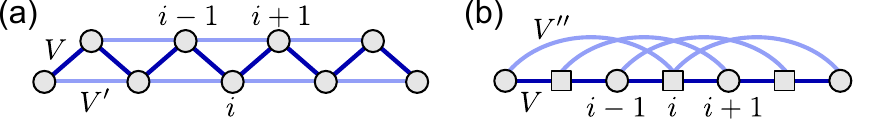}
	\caption{Schematic of the lattice configuration for engineering the nonlocal interaction $V_{ij}$. (a) The ladder-type lattice allows one to tune the NNN interaction $V^\prime$. (b) The dual-species scheme can suppress the NNN interaction while maintaining the NN interaction $V$ and the NNNN interaction $V^{\prime\prime}$.}
	\label{fig:fig4}
\end{figure}

In addition to the largest symmetry sector, we also investigate the ergodicity of the fully connected magnon sector with $n_\mathrm{NN}=0$. In this special sector, $H_\mathrm{eff}$ can be mapped to the constrained XXZ Hamiltonian \cite{pozsgay2021integrable,bastianello2022fragmentation}, which is integrable even in the presence of the three-body Ising interaction term. As a result, even for an intermediate hopping ratio $\xi\sim 2$, with which a general Krylov subspace thermalizes, e.g., the one considered in Fig.~\ref{fig:fig3}(b), eigenstates in the magnon sector still violate the ETH. This is verified by the broadened eigenstate entanglement entropy shown in Fig.~\ref{fig:fig3}(g), where we consider a $1/3$-filling magnon sector generated by the $\mathbb{Z}_3$ state $|{\bullet}{\circ} {\circ} {\bullet}{\circ} {\circ} \cdots\rangle$ with a system size $L=24$. The corresponding level spacing also follows a Poisson distribution [see Fig.~\ref{fig:fig3}(h)]. Note that the magnon sector has spatial inversion symmetry, so we only consider the inversion-symmetric states in these calculations. Similar to isolated magnons, we have verified that the fully connected hole sector also possesses this emergent integrability.

\begin{figure*}
	\centering
	\includegraphics[width=\linewidth]{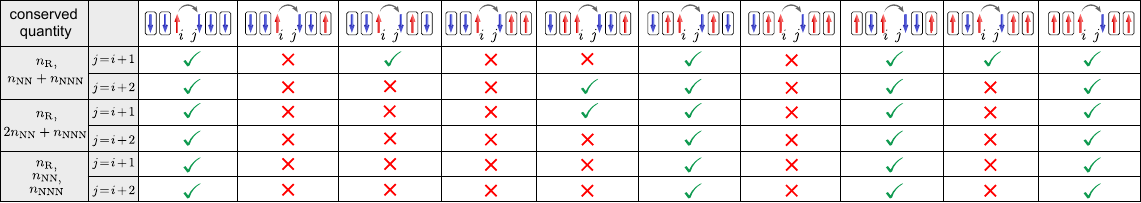}
	\caption{Kinetic constraints under different symmetries. The symmetry quantum numbers are given by $\{n_\mathrm{R},n_\mathrm{NN}+n_\mathrm{NNN}\}$, $\{n_\mathrm{R},2n_\mathrm{NN}+n_\mathrm{NNN}\}$, and $\{n_\mathrm{R},n_\mathrm{NN},n_\mathrm{NNN}\}$ for $V^\prime\approx V$, $V^\prime\approx V/2$, and other parameter regimes, respectively. The label of the chart indicates the NN ($j=i+1$) and the NNN ($j=i+2$) spin exchanges $(\sigma_i^+\sigma_j^-+ \sigma_i^-\sigma_j^+)$ with 10 irreducible spin states at sites $i-2$, $i-1$, $j+1$, $j+2$. The presence and absence of each term are indicated by ``$\checkmark$'' and ``$\times$'', respectively.}
	\label{fig:fig5}
\end{figure*}
\section{Nonlocal Rydberg Interactions}\label{sec:sec3}
Since a realistic Rydberg system usually features a nonlocal interaction, e.g., the long-range vdW interaction $V_{ij}=V/(i-j)^6$, we study in this section how the nonlocal interaction term $V_\mathrm{Ryd}=\sum_{i<j} V_{ij} {n}_i {n}_{j}$ modifies the properties of the HSF discussed in Sec.~\ref{sec:sec2}.

Similar to Sec.~\ref{subsec:sec2a}, we can perform the SW transformation and derive an effective Hamiltonian in the large detuning regime. For a nonlocal Rydberg interaction $V_{ij}$, the generator can be formally expressed as
\begin{equation}
\mathcal{S}=\mathrm{i}\frac{\Omega}{2}\sum_i\frac{{\sigma}_i^y}{\Delta + \sum_{j(\neq i)}V_{ij}{n}_j}.
\end{equation}
Note that the denominator of the generator is operator-valued, so an additional expansion of $\mathcal{S}$ is needed if one tries to obtain an explicit effective Hamiltonian through Eq.~\eqref{eq:eq3}. As in Eq.~(\ref{eq:eq4}), such expansion can be made by considering all possible spin configurations near a given site, for which the number of terms depends on the range of the interaction considered, i.e., $4^l$ terms are needed if we consider interactions up to a distance of $l$ lattice sites ($V_{i,i+l}$). The expansion can also be carried out in an alternative manner: expanding $\mathcal{S}$ in orders of the Rydberg excitation number that can influence the energy cost of the spin flip occurring at a given site, i.e.,
\begin{widetext}
\begin{equation}
	{\mathcal{S}} =  \frac{\mathrm{i}}{2}\sum_i \frac{\Omega}{\Delta}{\sigma}_i^y + \frac{\mathrm{i}}{2}\sum_{i\neq j}\left(\frac{\Omega}{\Delta+V_{ij}}-\frac{\Omega}{\Delta}\right){\sigma}_i^y{n}_j
	+\frac{\mathrm{i}}{4}\sum_{i\neq j\neq k}\left(\frac{\Omega}{\Delta+V_{ij}+V_{ik}}-\frac{\Omega}{\Delta+V_{ij}}-\frac{\Omega}{\Delta+V_{ik}}+\frac{\Omega}{\Delta}\right){\sigma}_i^y{n}_j{n}_k +\cdots,
\end{equation}
\end{widetext}
in which the number of terms becomes $2l+1$. While the above treatment works for a generic case, it is usually complicated to get an explicit Hamiltonian. Therefore, we will illustrate the nonlocal effect via special cases here.

\subsection{Strong nonlocal interaction: enriched kinetic constraint and fragmentation}\label{subsec:sec3a}
The inclusion of a strong nonlocal interaction can modify not only the effective Hamiltonian but also the fragmentation structure of the Hilbert space. As an example, we consider a minimal extension with $V_{ij}$ kept to the next-nearest-neighbor (NNN) interaction $V^\prime = V_{i,i+2}$ comparable to the NN interaction $V=V_{i,i+1}$. This can be achieved by adjusting the tilting angle of the staggered lattice realization shown in Fig.~\ref{fig:fig4}(a). The zeroth order Hamiltonian for such a system is given by
\begin{equation}
	{H}_\mathrm{0} =  \Delta n_\mathrm{R} + V n_\mathrm{NN} + V^\prime n_\mathrm{NNN},
\end{equation}
where $n_\mathrm{NNN}=\sum_{i} {n}_i {n}_{i+2}$ counts the total number of NNN Rydberg dimers.

Due to the conservation of the total number of Rydberg excitations $n_\mathrm{R}$, the off-diagonal part of the effective Hamiltonian is governed by the flip-flop terms as in Eq.~\eqref{eq:eq5}, albeit with an extended distance. The strengths of these spin-exchange interactions are of the same order as $J_P$ and $J_Q$ defined in Eq.~\eqref{eq:eq6}, and are hence much smaller than both the NN interaction $V$ and the NNN interaction $V^\prime$. Similar to the previous section, these strong diagonal interactions can give rise to distinct kinetic constraints. To determine them, we should consider the energy conservation imposed by the zeroth order Hamiltonian, i.e., only flip-flops that conserve the energy set by $H_0$ are allowed. In general, these constrained flip-flops can be described by
\begin{align}
	\mathcal{H}_\mathrm{off} &= \sum_{n,i} J_n\mathcal{P}_{i-2,i-1,i+2,i+3}^{[n]} \left(\sigma_{i}^+ \sigma_{i+1}^- + \sigma_{i}^-\sigma_{i+1}^+\right) \nonumber \\
&+\sum_{n,i} J_n^\prime\mathcal{P}_{i-2,i-1,i+1,i+3,i+4}^{\prime[n]} \left(\sigma_{i}^+ \sigma_{i+2}^- + \sigma_{i}^-\sigma_{i+2}^+\right),\label{eq:eq14}
\end{align} 
where $\mathcal{P}_{i-2,i-1,i+2,i+3}^{[n]}$ and $\mathcal{P}_{i-2,i-1,i+1,i+3,i+4}^{\prime[n]}$ project nearby sites of the $i$-th atom onto distinct spin configurations labeled by the index $n$, and the constraint can be described by a specific vanishing hopping strength $J_n=0$ or $J_n^\prime=0$. As we will discuss below, the form of the constraint is determined by the ratio $V^\prime/V$ and characterized by distinct emergent symmetry quantum numbers. The results are summarized in Fig.~\ref{fig:fig5}, where ``$\checkmark$'' and ``$\times$'' denote the presence and absence of the corresponding term in Eq.~\eqref{eq:eq14}, respectively.

\begin{figure}
	\centering
	\includegraphics[width=\linewidth]{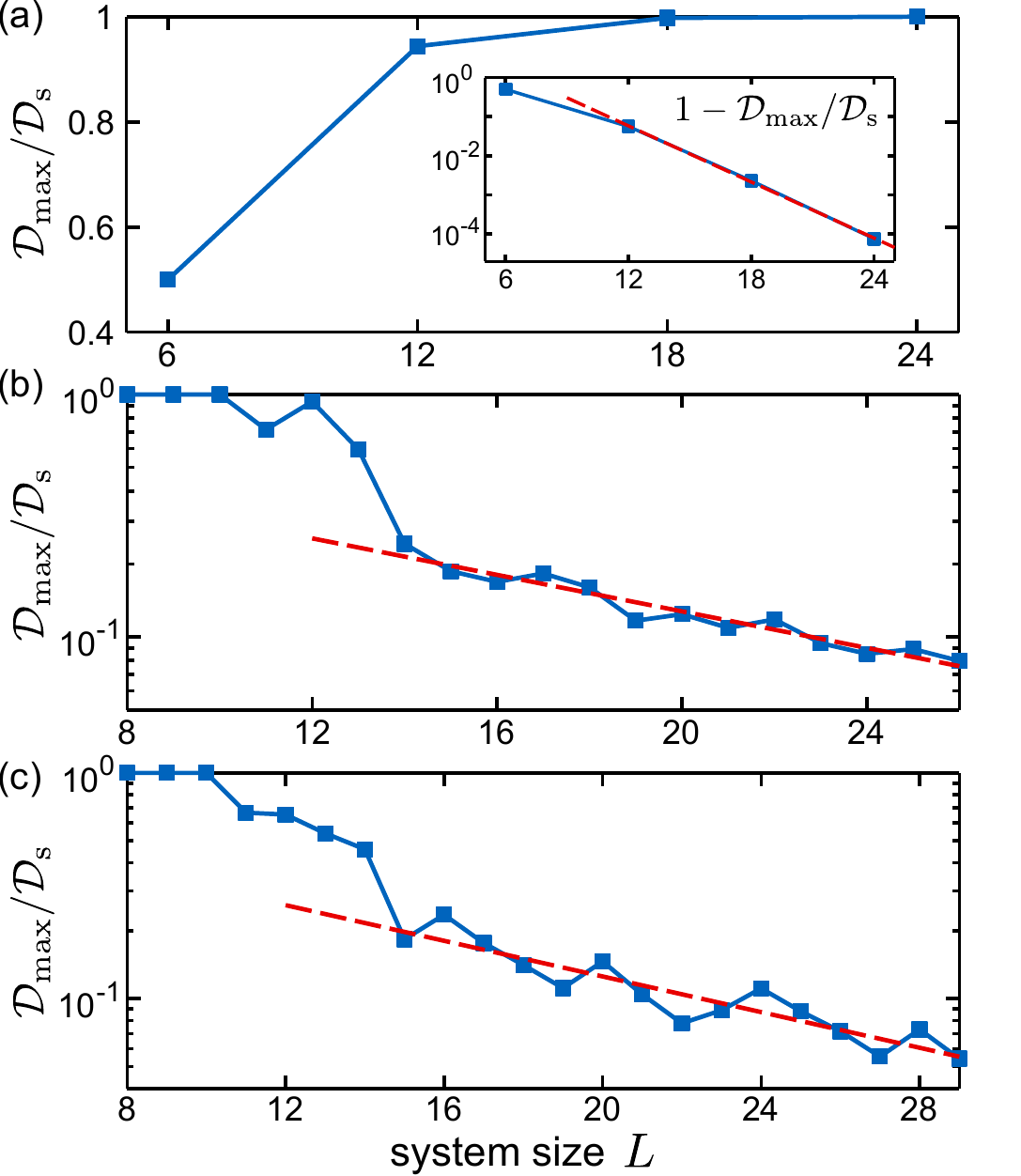}
	\caption{Characterization of the HSF in different regimes. (a) The Hilbert space for symmetry quantum numbers $\{n_\mathrm{R}=3m+1,n_\mathrm{NN}+n_\mathrm{NNN}=3m+1\}$ exhibits weak HSF with $\mathcal{D}_\mathrm{max}/\mathcal{D}_\mathrm{S}\approx 1-43.1\times0.577^L$ based on data up to $L=24$. (b) The largest symmetry sector with conserved charges $\{n_\mathrm{R},2n_\mathrm{NN}+n_\mathrm{NNN}\}$ exhibits strong HSF with $\mathcal{D}_\mathrm{max}/\mathcal{D}_\mathrm{S}\approx0.722\times0.917^L$ based on data up to $L=26$. (c) The largest symmetry sector with conserved charges $\{n_\mathrm{R},n_\mathrm{NN},n_\mathrm{NNN}\}$ exhibits strong HSF with $\mathcal{D}_\mathrm{max}/\mathcal{D}_\mathrm{S}\approx0.773\times0.913^L$ based on data up to $L=29$. The dashed lines represent exponential fits to the data.}
	\label{fig:fig6}
\end{figure}
First, for $V^\prime\approx V$, or more precisely, $|V^\prime- V|\lesssim J_P,J_Q$, the kinetic constraints are depicted by the first two rows of Fig.~\ref{fig:fig5}, with which we can identify two conserved charges $n_\mathrm{R}$ and $n_\mathrm{NN}+n_\mathrm{NNN}$. In contrast to the NN interaction regime, the largest symmetry sector now becomes a fully connected one without fragmentation due to the NNN hopping process. Nevertheless, one can still identify a weakly fragmented Hilbert space for other symmetry sectors, in which the largest Krylov subspace approaches the entire sector in the thermodynamic limit. For example, for symmetry quantum numbers $n_\mathrm{R}=3m+1$ and $n_\mathrm{NN}+n_\mathrm{NNN}=3m+1$ in a system with $L=6m$ atoms, the dimension of the largest Krylov subspace $\mathcal{D}_\mathrm{max}$ approaches the dimension of the sector $\mathcal{D}_\mathrm{S}$ with $1-\mathcal{D}_\mathrm{max}/\mathcal{D}_\mathrm{S}\sim 0.577^L$ [see Fig.~\ref{fig:fig6}(a)]. With such a weak HSF, the system can weakly violate the ETH by forming quantum many-body scars.

When the NNN interaction is tuned close to half of the NN interaction, i.e., $|V^\prime- V/2|\lesssim J_P,J_Q$, the allowed spin-exchange terms are reduced (see the middle two rows of Fig.~\ref{fig:fig5}), and the symmetry quantum number $n_\mathrm{NN}+n_\mathrm{NNN}$ is replaced by $2n_\mathrm{NN}+n_\mathrm{NNN}$. The enhanced kinetic constraint then significantly modifies the ergodicity of the system. In particular, the strong fragmentation of the largest symmetry sector is restored, with $\mathcal{D}_\mathrm{max}/\mathcal{D}_\mathrm{S}\approx 0.722\times0.917^L$ based on data up to $L=26$ [see Fig.~\ref{fig:fig6}(b)].

For a generically strong NNN interaction away from the above regimes, i.e., $|V^\prime-V|,|V^\prime-V/2|\gg J_P,J_Q$, the total NN dimers $n_\mathrm{NN}$ and the NNN dimers $n_\mathrm{NNN}$ are individually conserved. After resolving all three conserved quantities, we find that the largest Hilbert space is still strongly fragmented, with $\mathcal{D}_\mathrm{max}/\mathcal{D}_\mathrm{S}\approx 0.773\times0.913^L$ based on data up to $L=29$ [see Fig.~\ref{fig:fig6}(c)].

Note that in the above discussion, the NNN interaction is considered to be strong when compared to the hopping strength ($V^\prime\gg J_P,J_Q$), but not necessarily with respect to the detuning. If $V$ and $V^\prime$ are much larger than the detuning $\Delta$, most of the terms in Eq.~\eqref{eq:eq14} are significantly suppressed, which causes the SBL discussed in Sec.~\ref{subsec:sec2b}.

\begin{figure}[b]
	\centering
	\includegraphics[width=\linewidth]{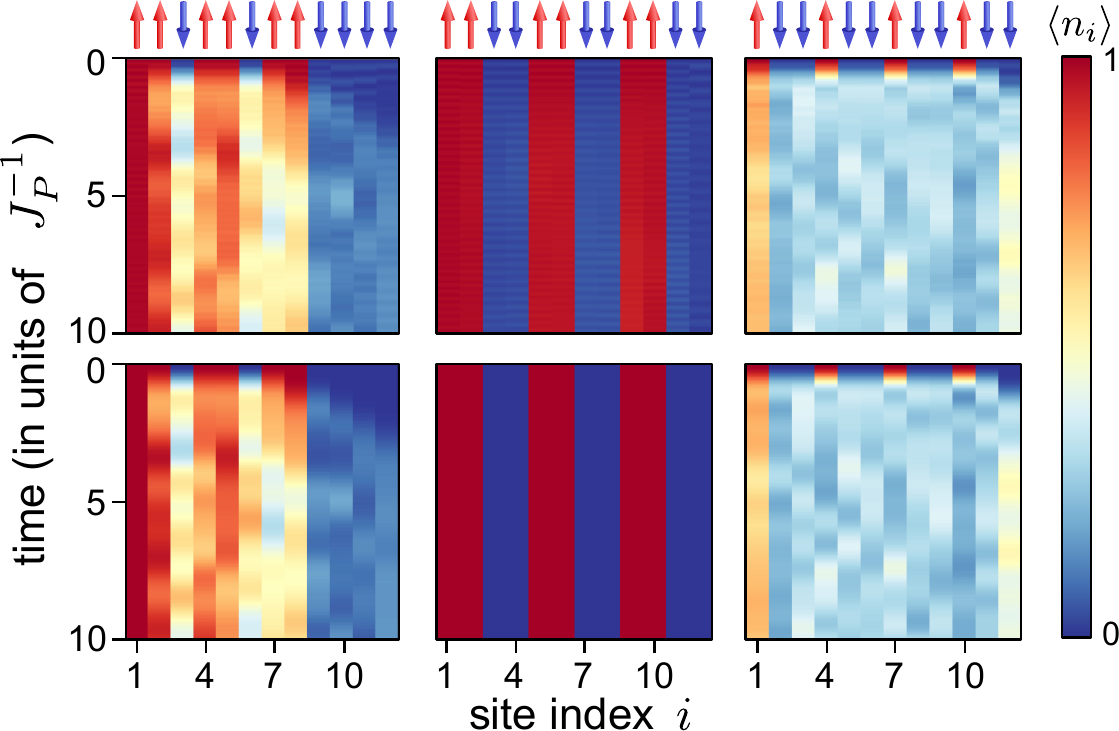}
	\caption{Quench dynamics for different indicated initial states in a Rydberg chain subjected to a nonlocal interaction. The upper and the lower panels show the evolution of the local Rydberg density $\langle n_{i} \rangle$ governed by the exact and the effective Hamiltonian, respectively. The parameters are $\Delta/\Omega=5$ and $V/\Delta=0.2$, and the power-law decaying vdW interaction $V_{ij}=V/|i-j|^6$ is retained up to $|i-j|=3$.}
	\label{fig:fig7}
\end{figure}
\subsection{Weak nonlocal interaction: integrability breaking}\label{subsec:sec3b}
We then consider the effect of a weak nonlocal interaction $V^\prime$ smaller or on the same order as the hopping strength $J_P,J_Q$. In this regime, the nonlocal term in $H_0$ is too weak to impose the kinetic constraint, such that the total number of NN Rydberg dimers $n_\mathrm{NN}$ is itself conserved as in the NN interaction regime. Moreover, the NNN flip-flop term $(\sigma_i^+\sigma_{i+2}^-+\sigma_i^-\sigma_{i+2}^+)$ in Eq.~\eqref{eq:eq14} can be safely dropped since its strength is two orders of magnitude smaller than the NN spin exchange $(\sigma_i^+\sigma_{i+1}^-+ \sigma_i^-\sigma_{i+1}^+)$. With these ingredients, the connectivity of the Hilbert space becomes exactly the same as the folded XXZ model discussed in the previous section. In fact, one can get the effective Hamiltonian by simply adding the term $V^\prime n_\mathrm{NNN}$ to $H_\mathrm{eff}$ derived for the NN interaction. This treatment can be generalized to a universal nonlocal interaction $V_{ij}$ satisfying $V_{|i-j|\geq 2}\lesssim J_P,J_Q$, i.e., the effective dynamics is governed by $H_\mathrm{eff}+\sum_{j-i\geq 2}V_{ij}{n}_i {n}_{j}$ with $H_\mathrm{eff}$ given by Eq.~(\ref{eq:eq7}). In Fig.~\ref{fig:fig7}, we show the quench dynamics for a realistic long-range interaction with the same initial states considered in Fig.~\ref{fig:fig2}(a). The nice agreement between the exact simulation and the above effective model confirms that a weak nonlocal interaction preserves the HSF and modifies only details of the dynamics.
\begin{figure}
	\centering
	\includegraphics[width=\linewidth]{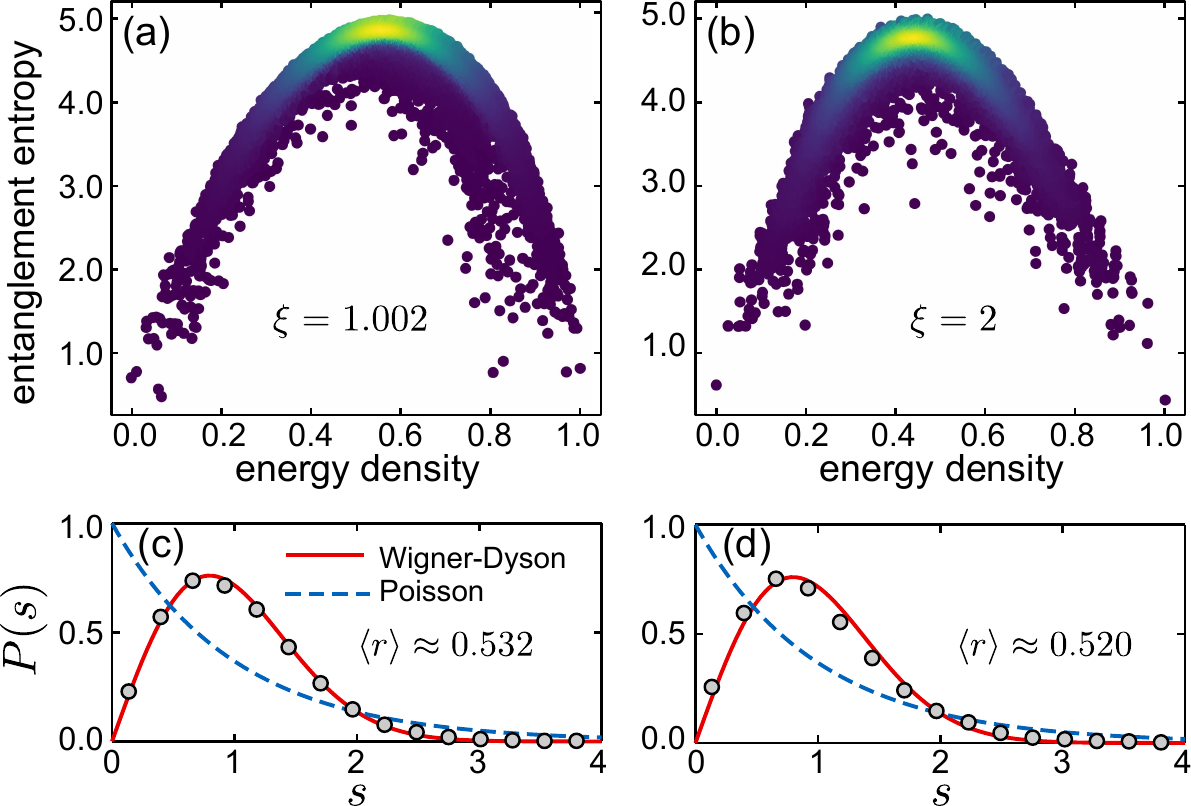}
	\caption{Half-chain von Neumann entanglement entropy of eigenstates and level spacing statistics under a weak nonlocal interaction. (a) Breaking of the approximate integrability for the largest Krylov subspace considered in Fig.~\ref{fig:fig3}(a). The nonlocal interaction is retained up to the next-next-nearest neighbor with $V^\prime=V/64$ and $V^{\prime\prime}=V/729$. (b) Breaking of the emergent integrability for the magnon sector considered in Fig.~\ref{fig:fig3}(g). The nonlocal interaction is retained up to the next-next-nearest neighbor with $V^\prime=0$ and $V^{\prime\prime}=V/729$.}
	\label{fig:fig8}
\end{figure}

We will focus on the weak nonlocal interaction regime in the rest of this article, as it is typical in a linear Rydberg chain where $V_{|i-j|\geq 2}\leq V/64$. While preserving the HSF, the long-range diagonal term $\sum_{j-i\geq 2}V_{ij}{n}_i {n}_{j}$ can still modify the ergodicity of the Krylov subspace. As discussed in Sec.~\ref{subsec:sec2a}, the system features an approximate integrability in the short-range interacting regime when the three-body interaction $I$ is suppressed by a $\xi$ close to unity. Here, the nonlocal interaction serves as an alternative integrability-breaking term, e.g., a small NNN interaction $V^\prime$ suffices to break such an approximate integrability \cite{pandey2020adiabatic,PhysRevResearch.5.043019}. Comparing Fig.~\ref{fig:fig8}(a) with Fig.~\ref{fig:fig3}(a), we note that the eigenstate entanglement entropy gets compressed by the nonlocal interaction and approaches the ergodic phase. The corresponding level spacing statistics also switch from the Poisson distribution [Fig.~\ref{fig:fig3}(d)] to the Wigner-Dyson distribution  [Fig.~\ref{fig:fig8}(c)].

The situation is different for the fully connected magnon sector, whose emergent integrability originates from the constrained XXZ model. This model remains Bethe Ansatz solvable for $V^\prime\neq 0$, such that longer-range interactions are required for the integrability breaking. As we show in Figs.~\ref{fig:fig8}(b) and \ref{fig:fig8}(d), by including a small next-next-nearest-neighbor (NNNN) interaction $V^{\prime\prime}=V_{i,i+3}$, the otherwise integrable subspace considered in Figs.~\ref{fig:fig3}(g) and \ref{fig:fig3}(h) begins to thermalize. Here, the ETH behavior is more pronounced with increasing $V^{\prime\prime}$, which inevitably leads to a larger $V^{\prime}$ that can modify the kinetic constraint as in Sec.~\ref{subsec:sec2a}. Note that the modified constraint can restore the integrability of the magnon sector by restricting the dynamics to a subspace described by the extended constrained XXZ model, where magnons are separated by more than two lattice sites \cite{alcaraz1999exactly,karnaukhov2002one}. Therefore, to clearly probe the thermalization in a finite-size system, one can implement a dual-species scheme [see Fig.~\ref{fig:fig4}(b)], in which the intra-species interaction $V^\prime$ and the inter-species interaction $V^{\prime\prime}$ can be independently tuned \cite{young2021asymmetric,singh2023mid}.

\section{Position disorder induced many-body localization}\label{sec:sec4}
In a state-of-the-art Rydberg array setup, the center-of-mass motion of the trapped atom is thermal. While motion of atoms during the spin dynamics can be neglected at sufficiently low temperatures (so-called ``frozen-gas'' limit), their initial positions are uncertain for each experimental realization. Such positional uncertainties can be treated as a static disorder, which can induce localization in the anti-blockade Rydberg chain \cite{marcuzzi2017facilitation,ostmann2019localization} as well as the Rydberg Heisenberg chain \cite{braemer2022pair}. Here, we investigate how this type of disorder modifies the ergodicity of the system studied in previous sections, and in particular, possible existence of the ergodic-MBL phase transition.
\begin{figure}[b]
	\centering
	\includegraphics[width=\linewidth]{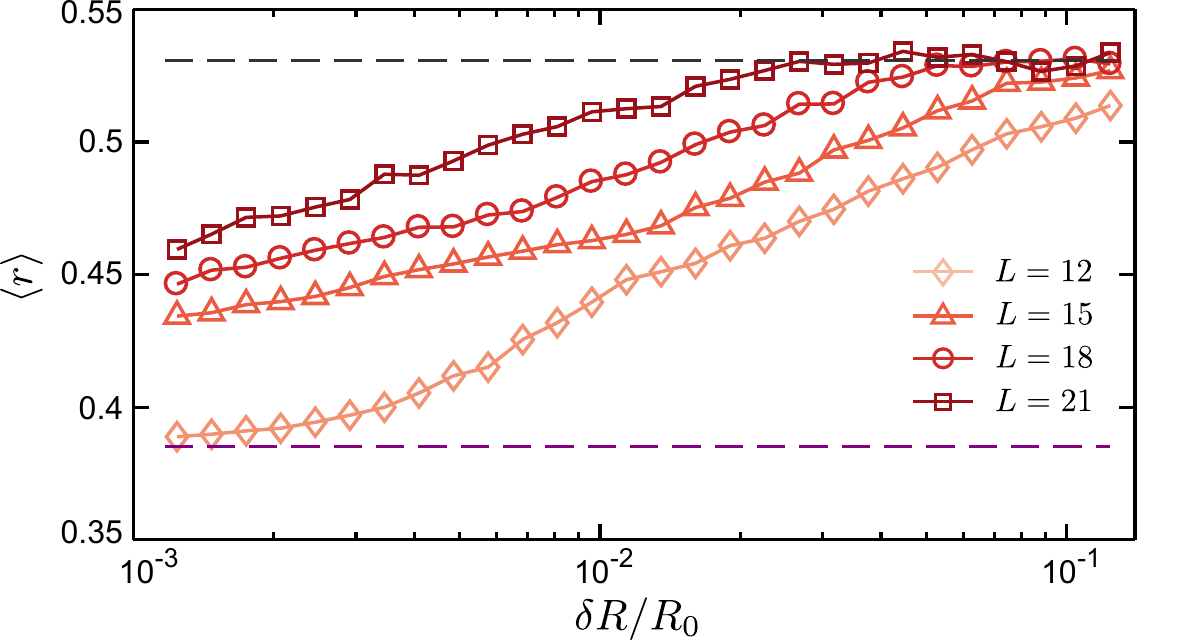}
	\caption{Mean level spacing ratio $\langle r\rangle$ as a function of the disorder strength $\delta R/R_0$ for the $1/3$-filling magnon sector. The results are averaged over 50 eigenstates closest to the middle of the spectrum and over 1000 disorder realizations (500 realizations for the largest system size $L=21$). The dashed curves indicate the two limits $\langle r\rangle_\mathrm{Poisson}$ and $\langle r\rangle_\mathrm{GOE}$ for the localized phase and the thermalized-ergodic phase, respectively. The parameters are given by $\Delta/\Omega=4$, $V/\Delta=0.2$, $V^\prime=V/64$, and $V^{\prime\prime}=V/729$.}
	\label{fig:fig9}
\end{figure}

\begin{figure*}
	\centering
	\includegraphics[width=\linewidth]{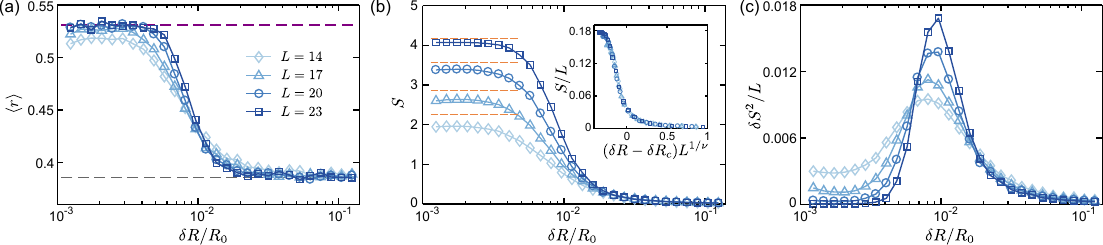}
	\caption{Characterization of the ergodic-MBL phase transition for the hole sector. (a) Mean level-spacing ratio $\langle r\rangle$ as a function of the disorder strength $\delta R/R_0$. The dashed curves indicate the two limits $\langle r\rangle_\mathrm{Poisson}$ and $\langle r\rangle_\mathrm{GOE}$. (b) Half-chain entanglement entropy $S$ as a function of $\delta R/R_0$. The dashed curves indicate the rescaled Page values $0.9S_\mathrm{Page}$, where $S_\mathrm{Page}=\sum_{k=n+1}^{mn}1/k-(m-1)/(2n)$ denotes the Page value for the considered sector with $m$ ($n$) being the minimal (maximal) Hilbert-space dimension of the partition. The inset displays the finite-size scaling result. (c) Variance of the entanglement entropy $\delta S^2$ as a function of $\delta R/R_0$. The results are averaged over 50 eigenstates closest to the middle of the spectrum and over 1000 disorder realizations (500 realizations for the largest system size $L=23$). The parameters are the same as in Fig.~\ref{fig:fig9}.}
	\label{fig:fig10}
\end{figure*}
In a perfectly ordered linear chain, the $j$-th atom is located at $jR_0$ with $R_0$ the lattice constant. In the presence of thermal fluctuations, atomic positions are modified as $R_j = jR_0 + \delta R_j$ with $\delta R_j$ a random number following a Gaussian distribution, whose variance depends on the depth of the trap and the temperature of the atom \cite{de2018analysis}. For simplicity, we assume here a uniform distribution of $\delta R_j\in[-\delta R/2,\delta R/2]$. The fluctuations in atomic positions then contribute correlated disorders $\delta V_{ij}$ in the vdW interaction $V_{ij}=C_6/(R_{i}-R_j)^6 = V/|i-j|^6 +\delta V_{ij}$. To get the disordered effective Hamiltonian, we need to perform the SW transformation with an inhomogeneous NN interaction $V_{i,i+1}$, which admixes the disorder to both the diagonal and the off-diagonal terms in Eq.~\eqref{eq:eq7}. When considering a weak nonlocal interaction as in Sec.~\ref{subsec:sec3b}, the diagonal disorders $\sum_{j-i\geq 2}\delta V_{ij}{n}_i {n}_{j}$ should also be included.

An important feature in our scheme is that the position disorder acts very differently on the transport of magnon excitations and hole excitations. To illustrate this, we first consider the magnon sector with $1/3$ filling fraction generated by the root state $|{\bullet}{\circ}{\circ} {\bullet}{\circ}{\circ} \cdots {\bullet}{\circ}{\circ}\rangle$ and compute the averaged gap ratio $\langle r\rangle$ for eigenstates close to the middle of the spectrum. As shown in Fig.~\ref{fig:fig9}, the parameter $\langle r \rangle$ drifts towards $\langle r\rangle_\mathrm{GOE}$ with increasing system size for the entire range of disorder strengths considered here, suggesting a preserved thermal behavior. Note that the disorder strength cannot increase further, otherwise the effective Hamiltonian itself breaks down as $n_\mathrm{NN}$ is no longer conserved. For weak disorder, the system remains close to inversion-symmetric, which causes deviations from the GOE value in systems of finite size. These simulations thus indicate the absence of MBL for the magnon excitations, whose robustness against position disorder is enabled by the Fibonacci constraint $n_in_{i+1}=0$ that eliminates the dominant disorder term $D = \sum_{i}\delta V_{i,i+1}n_in_{i+1}$.

This leading disorder term, however, plays a significant role for the hole sector governed by the local constraint $(1-n_i)(1-n_{i+1})=0$, under which $D$ becomes an on-site disorder term $D=\sum_{i}(\delta V_{i-1,i}+\delta V_{i,i+1}){n}_i$, akin to the Anderson model \cite{anderson1958absence}. As a result, a disordered interaction strength $\delta V=6V(\delta R/R_0)$ larger than the hopping strength $J_Q$ is expected to slow down the motion of the holes. We then study the spectral property of the $2/3$-filling hole sector generated by the root state $|{\bullet}{\bullet}{\circ} {\bullet}{\bullet}{\circ} \cdots {\bullet}{\bullet}{\circ} {\bullet}{\bullet}\rangle$. Here, a weak disorder $\delta R\sim 0.001 R_0$ is sufficient to break the inversion symmetry of the hole sector of the same Hilbert-space dimension and make the level-spacing parameter $\langle r\rangle$ approach the GOE value [see Fig.~\ref{fig:fig10}(a)]. Furthermore, at a strong disorder $\delta R\sim 0.1 R_0$, $\langle r\rangle$ is close to the Poisson value $\langle r\rangle_\mathrm{Poisson}$, which is a strong indication of localization. In the intermediate region, a crossover from the thermal phase to the localized phase can be identified, which becomes sharper with increasing system size $L$. Here, the crossing point between curves corresponding to $L$ and $L-3$ atoms drifts towards a larger disorder strength with a slower speed as $L$ increases, suggesting a possible ergodic-MBL phase transition at a finite disorder strength in the thermodynamic limit.

To further confirm the existence of the ergodic-MBL phase transition and to identify the critical point $\delta R_c$, we investigate properties of the eigenstates close to the middle of the spectrum, including the half-chain entanglement entropy $S$ and its variance $\delta S^2$. In a thermal phase, these high-temperature states are highly entangled, satisfying the volume law $S\propto L$. In comparison, all eigenstates in the MBL phase are short-range entangled due to the extensive set of local integrals of motion \cite{imbrie2017local,rademaker2017many} and will instead exhibit an area-law entanglement. Near the critical point, the sudden change of the eigenstate property yields a very large $\delta S^2$, whose peak value scales superlinearly with $L$ for a finite system size.

These features can be clearly identified in our simulations, where the bipartite entanglement entropy $S$ switches from the volume law to the area law when increasing the disorder strength  [see Fig.~\ref{fig:fig10}(b)], and its variance $\delta S^2$ features a narrow peak near the critical disorder strength [see Fig.~\ref{fig:fig10}(c)]. Based on the entanglement entropy, we perform the finite-size scaling, in which $S$ calculated for different system sizes and disorder strengths is fitted to the functional form $f[(\delta R-\delta R_c)^{1/\nu}]$. As shown in the inset of Fig.~\ref{fig:fig10}(b), the data collapse nicely onto each other with $\delta R_c/R_0\approx0.013$, which yields a critical interaction disorder $\delta V\approx8.15J_Q$. The critical exponent $\nu\approx0.93$ is close to the value found for the MBL transition in a Heisenberg chain with uncorrelated disorders \cite{luitz2015many,khemani2017two}, suggesting that these phase transitions may belong to the same universality class.

For a general Krylov subspace, the propagation of magnons and holes are coupled with each other. In this case, a strong disorder is expected to localize the hole while thermalizing the magnon dynamics, in the same manner as the SBL regime \cite{li2017statistical,yarloo2019emergent}.

\section{Toward experimental observation}~\label{sec:sec5}
The rapid progresses in Rydberg tweezer array setups, including the individual addressability and the improved atomic coherence \cite{evered2023high}, have enabled quantum simulation of many-body dynamics with single-site resolution at an intermediately long time scale. In this section, we discuss how to probe the ergodicity of the fragmented Hilbert space through many-body quench dynamics in a realistic Rydberg setup with experimentally feasible parameters $\Omega =2\pi\times 5~$MHz, $\Delta = 4\Omega$, and $V=0.2\Delta$. For a coherence time $T\sim 30~\mu$s, one can achieve an interaction-to-decay ratio $J_P T \sim 10$, which is sufficient for observing signatures of the phenomena discussed in the previous sections, as we will show below. Since our scheme considers states beyond the blockade regime, the repulsive forces between NN Rydberg excitations could affect the long-time dynamics. To suppress such a motional decoherence, state-independent trapping \cite{barredo2020three} and cooling \cite{belyansky2019nondestructive} of the atoms can be applied.

\begin{figure}
	\centering
	\includegraphics[width=\linewidth]{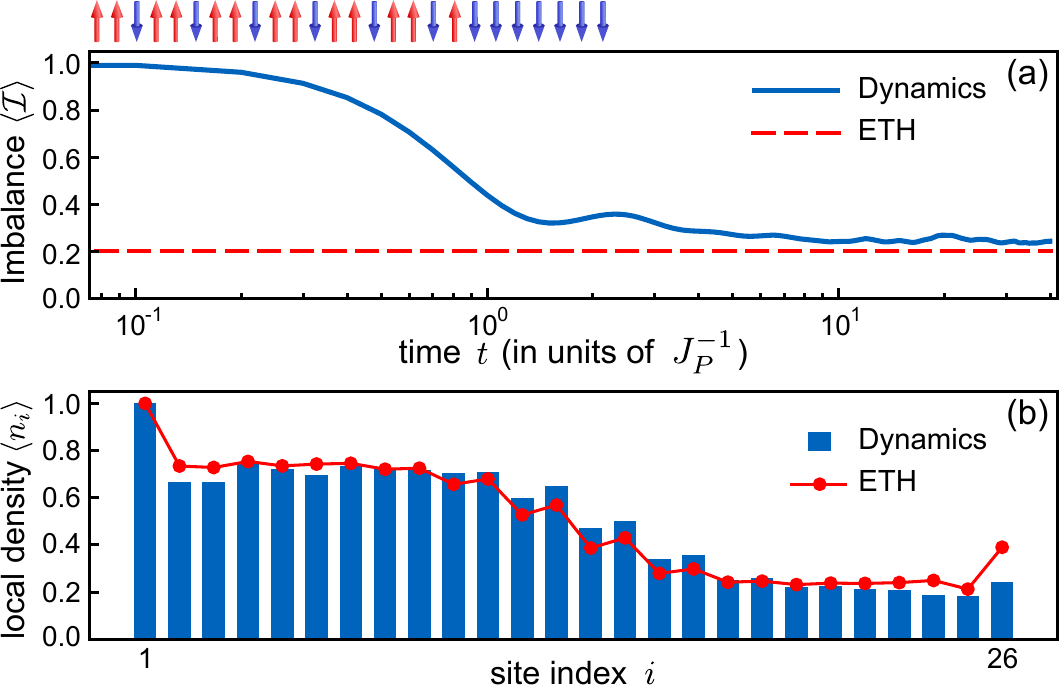}
	\caption{Characterization of the Krylov-restricted thermalization via quench dynamics. The initial spin configurations are indicated at the top of panel (a), which displays the evolution of the imbalance $\langle\mathcal{I}\rangle$. (b) shows the distribution of the local Rydberg density $\langle n_{i}\rangle$ at $t=40/J_P$. The label ``Dynamics'' [see the solid lines in (a) and the bars in (b)] represents the results obtained from the dynamics governed by the effective Hamiltonian. The label ``ETH'' [see the dashed lines in (a) and the connected dots in (b)] represents the results obtained by Eq.~\eqref{eq:ETH} with $\mathcal{N}=50$.}
	\label{fig:fig11}
\end{figure}

Note that the single-site-resolved fluorescence imaging permits extraction of any local observables \cite{browaeys2020many}, e.g., the Rydberg density $\langle n_i\rangle$, from the experiment. This allows us to characterize the thermalization in terms of the magnetization imbalance operator $\mathcal{I}$, which is defined as
\begin{equation}
\mathcal{I}(t)=\frac{\sum_i\langle  n_i(0)\rangle \sigma_i^{z}(t)}{2\sum_i\langle n_i(0)\rangle}-\frac{\sum_i\langle 1-n_i(0)\rangle \sigma_i^{z}(t)}{2\sum_i\langle 1-n_i(0)\rangle}
\end{equation}
in the Heisenberg picture \cite{morong2021observation}. For a product state consisting of up and down spins, the initial magnetization imbalance $\mathcal{I}(0)$ is unity, which is expected to decay to zero when the system is completely thermalized, losing the memory of its initial state. For example, in a standard XXZ model, an initial N\'eel state with zero total magnetization ($S_z=0$) is at infinite temperature and will rapidly thermalize with $\mathcal{I}(\infty)\approx0$. 

\begin{figure}[b]
	\centering
	\includegraphics[width=\linewidth]{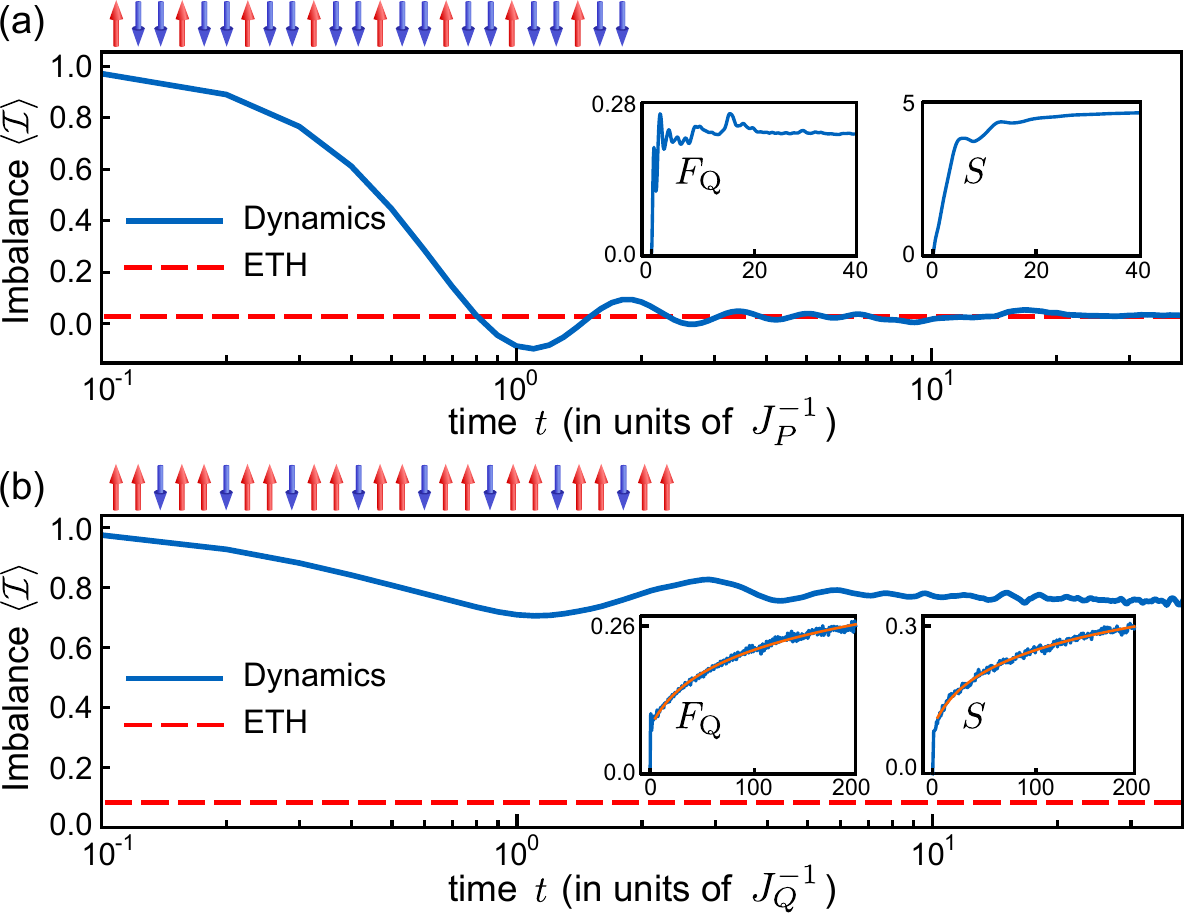}
	\caption{Characterization of the thermal phase and the MBL phase via quench dynamics. (a) and (b) show the evolution of the imbalance $\langle\mathcal{I}\rangle$ for the magnon sector with $L=24$ atoms and the hole sector with $L=26$ atoms, respectively. The initial spin states are illustrated at the top of each panel. The insets display the evolution of the quantum Fisher information $F_\mathrm{Q}$ and half-chain von Neumann entanglement entropy $S$, with the blue and the orange curves indicating the data and the logarithmic fit, respectively. The results are averaged over 100 disorder realizations with a disorder strength $\delta R/R_0 = 0.04$. The definitions of the labels ``Dynamics'' and ``ETH'' are the same as in Fig.~\ref{fig:fig11}.}
	\label{fig:fig12}
\end{figure}
Thermalization in a strongly fragmented system behaves rather differently. As described in Sec.~\ref{sec:sec2}, the zero-magnetization sector $S_z=0$ in our effective model features a strong HSF. When initializing the system with the $S_z=0$ state given by Eqs.~\eqref{eq:eq8} or \eqref{eq:eq9}, we find that the imbalance $\langle \mathcal{I}(t)\rangle$ will saturate at a finite value $\approx 0.2$ [see Fig.~\ref{fig:fig11}(a)], indicating that the system retains partial memory of the initial condition. This non-zero value can be predicted by the ETH if we restrict ourself to the Krylov subspace generated by the initial state, i.e., the expectation value of an observable $\mathcal{O}$ at equilibrium is
\begin{equation}
\langle \mathcal{O}\rangle_\mathrm{ETH} = \frac{1}{\mathcal{N}}\sum_{n=1}^{\mathcal{N}} \langle \left.\mathcal{O}\rangle_{n}\right|_{E_n\approx E},\label{eq:ETH}
\end{equation}
which is averaged over $\mathcal{N}$ eigenstates within the considered Krylov subspace, whose eigenenergies $E_n$ are closest to the energy $E$ of the initial state, i.e., ordered by $|E_n-E|$. Such Krylov-restricted thermalization can be further verified by probing the steady-state distribution of the local Rydberg density. As shown in Fig.~\ref{fig:fig11}(b), at sufficiently long times $t \gtrsim 10/J_P$, the profile of $\langle n_i\rangle$ resembles its initial distribution to a certain extent and can be nicely captured by the restricted ETH.

Evolution of the magnetization imbalance can also be used to distinguish the thermal phase and the MBL phase, where the local information is retained by the disorder in the latter case. According to Sec.~\ref{sec:sec4}, at a strong position disorder $\delta R>\delta R_c$, the magnon sector remains thermal, while the hole sector becomes many-body localized. Such a difference is confirmed by the dynamical evidence displayed in Fig.~\ref{fig:fig12}, where separated magnons thermalize at a steady-state imbalance close to zero [Fig.~\ref{fig:fig12}(a)], while isolated holes exhibit a very slow dynamics with $\langle\mathcal{I}(t)\rangle$ kept far away from the equilibration value predicted by the ETH [Fig.~\ref{fig:fig12}(b)].

To gain deeper insight in the thermalization and the localization, we can probe the evolution of the quantum Fisher information with respect to the imbalance operator \cite{smith2016many}: ${F}_\mathrm{Q}=4[\langle\mathcal{I}^2\rangle-\langle \mathcal{I}\rangle^2]$, which reveals the establishment of the second-order quantum correlations $\langle\sigma_i^z\sigma_j^z\rangle-\langle\sigma_i^z\rangle \langle\sigma_j^z\rangle$. As shown in the inset of Fig.~\ref{fig:fig12}, ${F}_\mathrm{Q}(t)$ grows very rapidly in the thermal phase and logarithmically in the MBL phase, in agreement with studies of a Heisenberg chain \cite{safavi2019quantum}. A similar behavior is observed for the half-chain entanglement entropy $S$ which carries information of the higher-order correlations: $S(t)$ grows ballistically and saturates in the thermal phase, while following a logarithmic growth in the MBL phase.

\section{Conclusion and Outlook}\label{sec:sec6}
In conclusion, we present a systematic study of the Hilbert space fragmentation in a Rydberg Ising chain under a detuned laser driving. With the SW transformation, we obtain the effective Hamiltonian of the system, which features an exotic kinetic constraint and can be mapped to a generalized, folded XXZ model with asymmetric hopping of magnon and hole excitations. We find that such an asymmetry allows us to tune the ergodicity of the system, as verified by a detailed study of the eigenstate entanglement entropy and the level-spacing statistics. The role of realistic nonlocal Rydberg interactions and atomic position disorders is also addressed. In the former case, we find that a strong nonlocal interaction can significantly modify the kinetic constraints and enrich the fragmentation behavior of the system, while a weak nonlocal interaction can break the emergent integrability of the system. In the latter case, when increasing the disorder strength in a reasonable range, we find that the dynamics of the magnon sector remains ergodic, while the hole sector exhibits a phase transition from thermal to MBL. We further show dynamical evidences for these phenomena, including evolution of the magnetization imbalance and growth of the quantum fluctuation in quench dynamics. Finally, we note that the ergodicity breaking based on the second-order effective Hamiltonian corresponds to the prethermal phase of the original model \cite{hart2022hilbert}. Nevertheless, one can still observe interesting non-thermal phenomena at intermediate time scales, which can be extended by using a weaker perturbation together with an increased coherence time.

Our scheme provides many possibilities for future exploration. For example, it would be interesting to investigate the ergodicity of the emergent Krylov subspaces induced by a strong nonlocal interaction, as demonstrated in Sec.~\ref{sec:sec3}. In particular, the measure-zero set in the regime $V^\prime\approx V$ allows one to construct quantum many-body scars with perfect revivals \cite{choi2019emergent,PhysRevB.106.205150}. Generalization of the scheme to higher dimensions may also enrich the fragmentation structure of the system \cite{yoshinaga2022emergence,hart2022hilbert} and give rise to interesting non-ergodic behavior. In the study of MBL, we mainly consider the magnon and the hole sector, while for a general Krylov subspace \cite{herviou2021many} containing both magnon and hole excitations, there might exist interesting partial localization phenomena. One can also apply an additional on-site disorder $\sum_i\Delta_i n_i$ through random laser detunings $\Delta_i$, with which the magnons can be localized as well. It is also interesting to investigate the stability of the observed MBL phase in the thermodynamic limit \cite{morningstar2022avalanches}. Moreover, the kinetically constrained dynamics offers a way to engineer lattice gauge models that can exhibit disorder-free localization \cite{smith2017disorder,homeier2023realistic,cheng2024emergent,sala2024disorder}.

In addition to the magnetization imbalance emphasized in the current paper, we recall that Rydberg arrays permit observation of many other important indicators for the ergodicity breaking and quantum chaos, such as the single-site-resolved spatial correlation, the out-of-time-ordered correlator \cite{fan2017out,blocher2022measuring,kastner2024ancilla}, the spectral form factor \cite{joshi2022probing}, and the entanglement entropy \cite{bluvstein2022quantum}. By applying laser fields of different frequencies, our scheme allows for the insertion of a synthetic gauge flux \cite{wu2022manipulating}, which can be used to study the Hilbert space fragmentation with broken time reversal symmetry in two dimensions. By including experimentally relevant dissipation terms \cite{kim2023realization}, one can also explore the behavior of the HSF in an open quantum system \cite{li2023hibert}.

\vspace{20pt}
{\it Note added}. After completion of this work, we became aware of a related study of Krylov-restricted thermalization with Rydberg atoms \cite{zhao2024observation}, where the Hilbert space is locally fragmented through Rydberg facilitation.

\begin{acknowledgments}
We thank A. Cooper, T. Pohl, L. You, X. Wu, J. Ahn, K. Kim, J. Li, Z. Wang and R. Gao for helpful discussions. F. Yang and K. M{\o}lmer acknowledge the support from the Carlsberg Foundation through the ``Semper Ardens'' Research Project QCooL. H. Yarloo, H.-C. Zhang, and A.~E.~B. Nielsen acknowledge the support from the Carlsberg Foundation under Grant No.~CF20-0658.
\end{acknowledgments}
	
\bibliography{Reference}
	
\end{document}